
\input aa.cmm
%
%
%
%
%
  \MAINTITLE={ Coordinated UV-optical observations of quasars: the
evolution of the Lyman absorption

  \FOOTNOTE{ Based on material collected with IUE and with the ESO-La Silla
telescopes.}
}
%
%
%
%
  \SUBTITLE={ ????? }
%
%
%
  \AUTHOR={ S. Cristiani @1,  E. Giallongo @2, L. M. Buson @3
C. Gouiffes @4, F. La Franca @{1} }
%
%
%
%
%
    \OFFPRINTS={ S. Cristiani }
%
%
%
  \INSTITUTE={
@1 Dipartimento di Astronomia della Universit\`a di Padova,
Vicolo dell' Osservatorio 5, I-35122 Padova, Italy
@2 Osservatorio Astronomico di Roma, Via dell' Osservatorio,
I-00040 Monteporzio, Italy
@3  Osservatorio Astronomico di Padova,
Vicolo dell' Osservatorio 5, I-35122 Padova, Italy
@4 European Southern Observatory, K. Schwarzschild Strasse 2,
D-8046 Garching, Germany}
%
%
  \DATE={ Received May 27, accepted Sep 24, 1992}
%
%
  \ABSTRACT={
The average flux decrement shortward the Ly$_{\alpha}$ emission, due to the
well-known ``forest'' of absorptions, has been measured in the spectra of 8
quasars. Quasi-simultaneous optical and IUE observations of the two low
redshift quasars PKS 0637--75 (z=0.654) and MC 1104+16 (z=0.632) have been
carried out, obtaining relatively high S/N, spectrophotometrically calibrated
data on their energy distribution from the rest frame H$_{\beta}$ to the Lyman
continuum. Six more quasars in the redshift range 2.5-3.4 have been observed
in the optical domain. For all the quasars the ``intrinsic'' continuum slope
and
normalization have been estimated longward the Ly$_{\alpha}$ emission and
extrapolated towards the Lyman continuum to measure the average depressions,
which have been compared with the model statistics of the Ly$_{\alpha}$
clouds. When all the known
classes of absorbers are taken into account with plausible values for their
equivalent width distribution and evolution, a good agreement is obtained with
the observations. The results for the observed continuum decrement at $z \sim
0.65$ are identical to those predicted by the evolution with redshift of the
number of Ly$_{\alpha}$  forest systems including the HST data and within
$2\sigma$ of the predicted value using the ``standard''
Ly$_{\alpha}$ evolution (as determined only at high z).
}
%
%
%
   \KEYWORDS={ Quasars: general -- Intergalactic medium}
%
%
%
  \THESAURUS={ 17.01.1; 09.05.1 }
%
%
\maketitle
%
%
%
%
%

\titlea{Introduction}

Considerable information about non-luminous extragalactic matter
has been gained by studying the absorption features in quasar spectra.
Particularly interesting in this respect is the issue of the evolution of the
absorbers responsible for the ``Ly$_\alpha$ forest systems''. It is generally
accepted that they are originated by intergalactic primordial clouds
(Sargent et al., 1980) and as such distinguished - although
recent observations tend to blur this difference - from the less numerous
``metal-line systems''.

Optical spectroscopy at high resolution available for a few QSOs at $z\sim 3$
shows a high number of hydrogen absorption lines whose
statistical properties can be represented, in terms of the number density of
hydrogen clouds, by a power-law in redshift with index $\sim 2$,
indicating a cosmological evolution of the absorbers.

At low resolution this ``forest'' of absorptions appears as
a steepening of the continuum slope in the region shortward the Ly$_\alpha$
emission  as compared to the slope defined on the red side of Ly$_\alpha$
(e.g. Steidel \& Sargent, 1987).
The average flux decrements $D_A$, between Ly$_\alpha$ and Ly$_\beta$
emissions,
and $D_B$, between Ly$_\beta$ and the Lyman limit have been introduced
to study the average properties of the absorption (Oke and Korycansky 1982,
Steidel and Sargent 1987). These quantities
are, within relatively large limits, independent of the instrumental
resolution, at least
as far as the {\it true continuum} can be reliably extrapolated from the
unabsorbed (red) region.

\begfigwid 15 cm
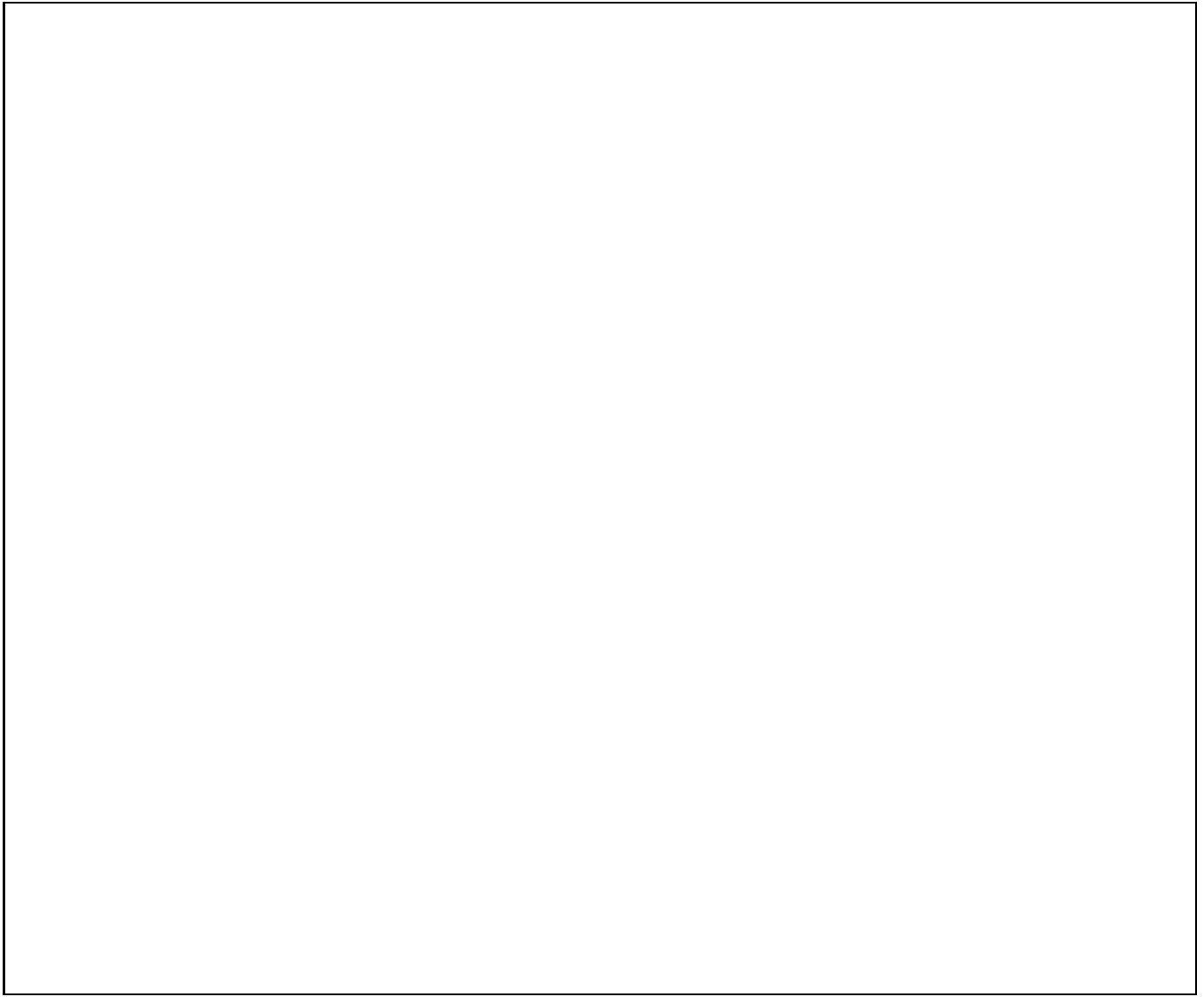
\figure{1}{IUE spectra of PKS 0637--75 and MC 1104+16, fluxes are in units of
10$^{-14}$ erg s$^{-1}$ cm$^{-2}$ \AA$^{-1}$ }
\endfig

Using the material available in the literature (Bechtold {\it et al.} ,
1984; O' Brien {\it et al.} , 1988; Oke and Korycansky, 1982; Schneider,
Schmidt and Gunn, 1989; Steidel and Sargent, 1987) together with
new optical data obtained by us at high redshifts, we have tried to
disentangle the evolutionary details of the Ly$_\alpha$ absorbers
(Giallongo \& Cristiani, 1990).
Excess absorption with respect to the $\gamma \sim 2$ evolution is
found both at high and low redshifts (as could be expected since
the evolutionary law has been normalized at z$\sim 3$ where the bulk
of the optical data has been obtained).
The presence of different populations of absorbers might be suggested
as well as a change in the physical conditions of the clouds. In any case
the slower evolution found at low redshifts resembles that of
metallic lines, perhaps indicating a link with the evolution of
normal galaxies.

The acquisition of new material is important especially at low redshifts,
because in this region the uncertainties are
still rather large and the constraints to the functional form of the
evolution of the absorbers imposed by the data at the lower extreme
of the observable redshift range can be particularly strong.
For this reason we have observed 2 bright QSOs (V $\sim 16$) with
z$\simeq 0.65$ (an especially convenient redshift for IUE observations
since the QSO Ly$_\alpha$ emission falls then in the region
of overlap between the SWP and the LWP cameras) to provide additional good
quality spectra needed to clarify the situation.
In addition, optical observations of 6 quasars have been carried out in
the optical domain, in order to improve the statistics and the quality of the
data at $z \simeq 2.5 - 3$.

\titlea{UV Observations}

Two consecutive IUE low--resolution long--wavelength spectra and one
short--wavelength spectrum of PKS~0637--75 have been obtained on
Dec. 25--26 1990 at VILSPA.
One SWP and one LWP spectrum secured during a subsequent run on
April 6--7 1991 allowed us a full coverage of the IUE spectral range
($\lambda\lambda$~1200$\div$3300~\AA) also for MC~1104+167.
Since the visual
magnitude of both objects (V$\sim$16) is beyond the imaging capabilities of the
IUE FES, a standard blind offset technique from nearby stars was adopted to put
the target into the slit.

\begtabfull
\tabcap{1}{ Log of used IUE observations.}
\halign{ # \hfill && #  \hfill \cr
\noalign{\medskip\hrule\medskip}
Object & Spectrum & Obs. Date & Exp.~T.~(m.) & ECC \cr
\noalign{\medskip\hrule\medskip}
PKS~0637--75  & SWP~10832 & 80~Dec~12 & 416 & --- \cr
PKS~0637--75  & LWP~13013 & 88~Apr~10 & 120 & 335 \cr
PKS~0637--75  & LWP~13016 & 88~Apr~11 & 105 & 335 \cr
PKS~0637--75  & LWP~13022 & 88~Apr~12 & 120 & 338 \cr
PKS~0637--75  & LWP~19471 & 90~Dec~25 & 180 & 331 \cr
PKS~0637--75  & LWP~19472 & 90~Dec~25 & 215 & 341 \cr
PKS~0637--75  & SWP~40461 & 90~Dec~26 & 404 & 331 \cr
MC~1104+167   & SWP~41309 & 91~Apr~06 & 400 & 352 \cr
MC~1104+167   & LWP~20094 & 91~Apr~07 & 360 & 452 \cr
\noalign{\medskip\hrule\medskip}
\noalign{\medskip\hrule\medskip}}\endtab

The exposure times spanned from 3 to 7 hours giving a
S/N ratio $10~\div~15$ in the best exposed regions. Taking into
account that PKS~0637--75 did not show significant UV variability over a
decade, four pre--existing lower--quality spectra collected at GSFC between
1980 and 1988 have been included in our analysis. Their flux levels have been
renormalized to the level observed in Dec 1990 and a weighted mean of all the
data has then been carried out.
A previous SWP spectrum of
MC~1104+167 does not contain signal and was discarded. A complete log of used
observations is reported in Table 1.  This table also includes, when
available, the VILSPA Exposure Classification Code (ECC) which gives an
approximate measure of the image quality.

\begfigwid 14 cm
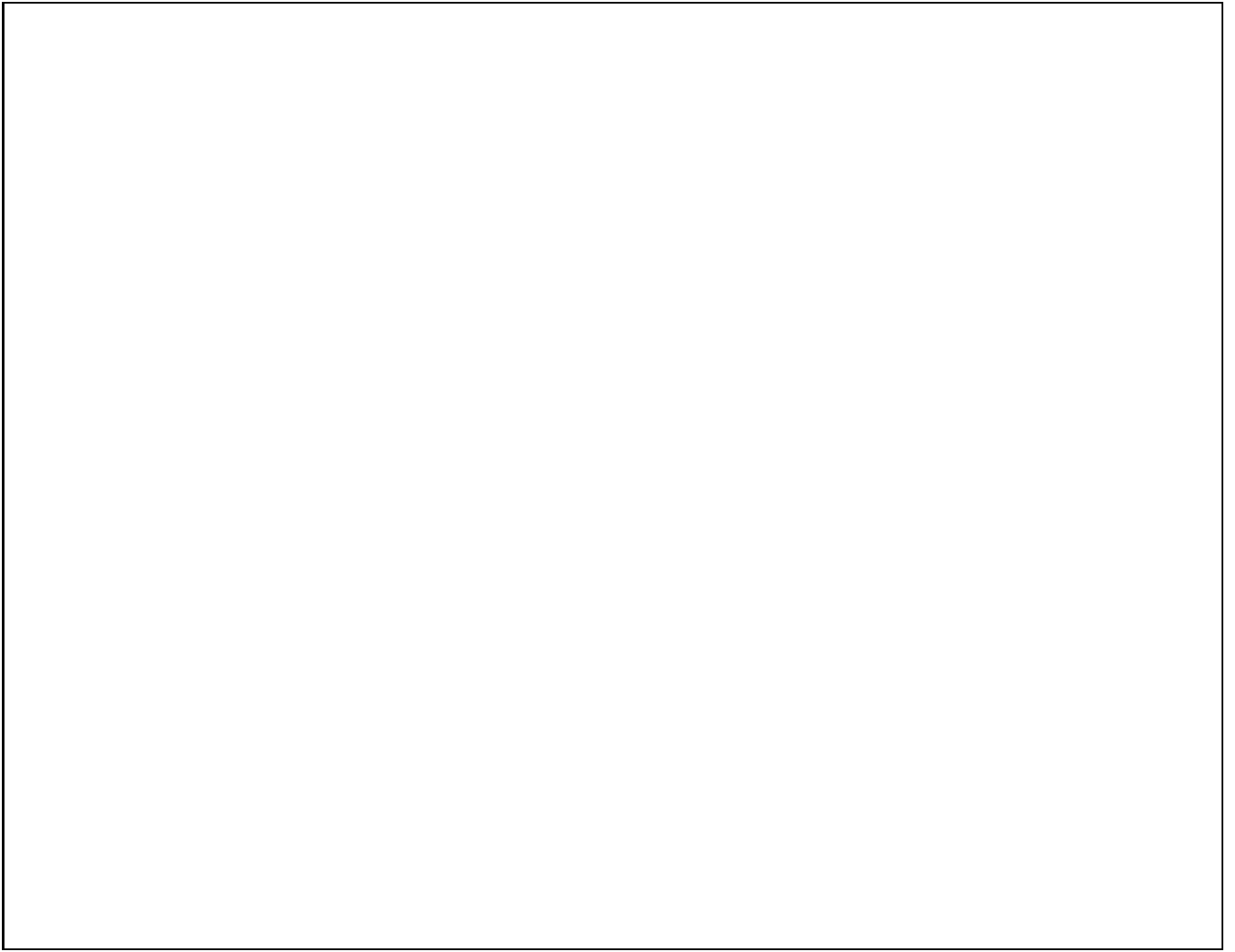
\figure{2a}{Optical spectra of Q0053--4029, Q0126--4050, Q0249+0222,
Q0254--0137, relative fluxes are in units of
10$^{-16}$ erg s$^{-1}$ cm$^{-2}$ \AA$^{-1}$ }
\endfig

Both our spectra and previous data have been consistently re--extracted from
the so--called {\it line--by--line} spectrum provided by the standard IUESIPS
processing. These spectra are already wavelength calibrated and photometrically
linearized. Our subsequent extraction includes removal of cosmic ray hits and
camera reseau marks as well as subtraction of a smoothed nearby background
giving a time--integrated net spectrum. The final absolute flux calibrated
spectrum is obtained by multiplying by the proper calibration function and
dividing by the exposure time. Since our data have been processed with the
current version of IUESIPS, the present LWP ITF2 calibration function
(Cassatella {\it et al.} 1988) has been adopted for long--wavelength spectra.
The short--wavelength data have been calibrated by means of the standard SWP
function (Holm {\it et al.} 1982). A test extraction including the signal
weighting on the basis of the observed PSF did not improve significantly our
final S/N ratio.

Both SWP and LWP data have been finally corrected for
the time--dependent camera sensitivity degradation as proposed by Bohlin and
Grillmair (1988) and Teays and Garhart (1990) respectively. The estimated
absolute flux error does not exceed 0.05 dex in both cameras.
The IUE spectra of both quasars are shown in Fig. 1.

\titlea{Optical Observations}

\begtabfullwid
\tabcap{2}{Journal of the optical spectroscopic observations}
\halign{ # \hfill & \hfill # \hfill && \hfill #  \hfill \cr
\noalign{\medskip\hrule\medskip}
Name & R.A.& Dec. & Date & Telescope & Spectrograph &
Slit width & Resolution & Sp. Range \cr
& (1950.0) & (1950.0) & & & & (arcsec) & (\AA) & (\AA) \cr
\noalign{\medskip\hrule\medskip}
Q0053--4029 & 00 53 51.5 & --40 29 30 & 02 Sep 1987 & 3.6m & EFOSC1 &
2.0 & 15 & 3750-7000 \cr
Q0126--4050 & 01 26 13.6 & --40 50 05 & 03 Sep 1989 & 2.2m & B.\&C. &
2.0 & 20 & 3400-8650 \cr
Q0249+0222 & 02 49 42.4 & +02 22 57 & 30 Nov 1989 & 3.6m & EFOSC1 &
2.0 & 15 & 3700-7000 \cr
Q0254--0137 & 02 54 07.9 & --01 37 49 & 30 Nov 1989 & 3.6m & EFOSC1 &
2.0 & 15 & 3700-7000 \cr
Q0256--0000 & 02 56 31.8 & --00 00 29 & 02 Jan 1984 & 3.6m & B.\&C. &
3.0 & 20 & 4350-8320 \cr
Q0301--0035 & 03 01 07.7 & --00 35 01 & 17 Oct 1984 & 3.6m & B.\&C. &
2.0 & 20 & 3700-8750 \cr
PKS0637--75 & 06 37 23.3 & --75 13 38 & 20 Dec 1990 & 3.6m & EFOSC1 &
1.5 & 15 & 3700-7000 \cr
PKS0637--75 & 06 37 23.3 & --75 13 38 & 19 Dec 1986 & 2.2m & B.\&C. &
6.0 & 30 & 3500-10000 \cr
MC 1104+16 & 11 04 36.7 & +16 44 17 & 13 Apr 1991 & 1.5m & B.\&C. &
2.0 & 22 & 4800-8800 \cr
\noalign{\medskip\hrule\medskip}
\noalign{\medskip\hrule\medskip}}\endtab

\begfigwid 14 cm
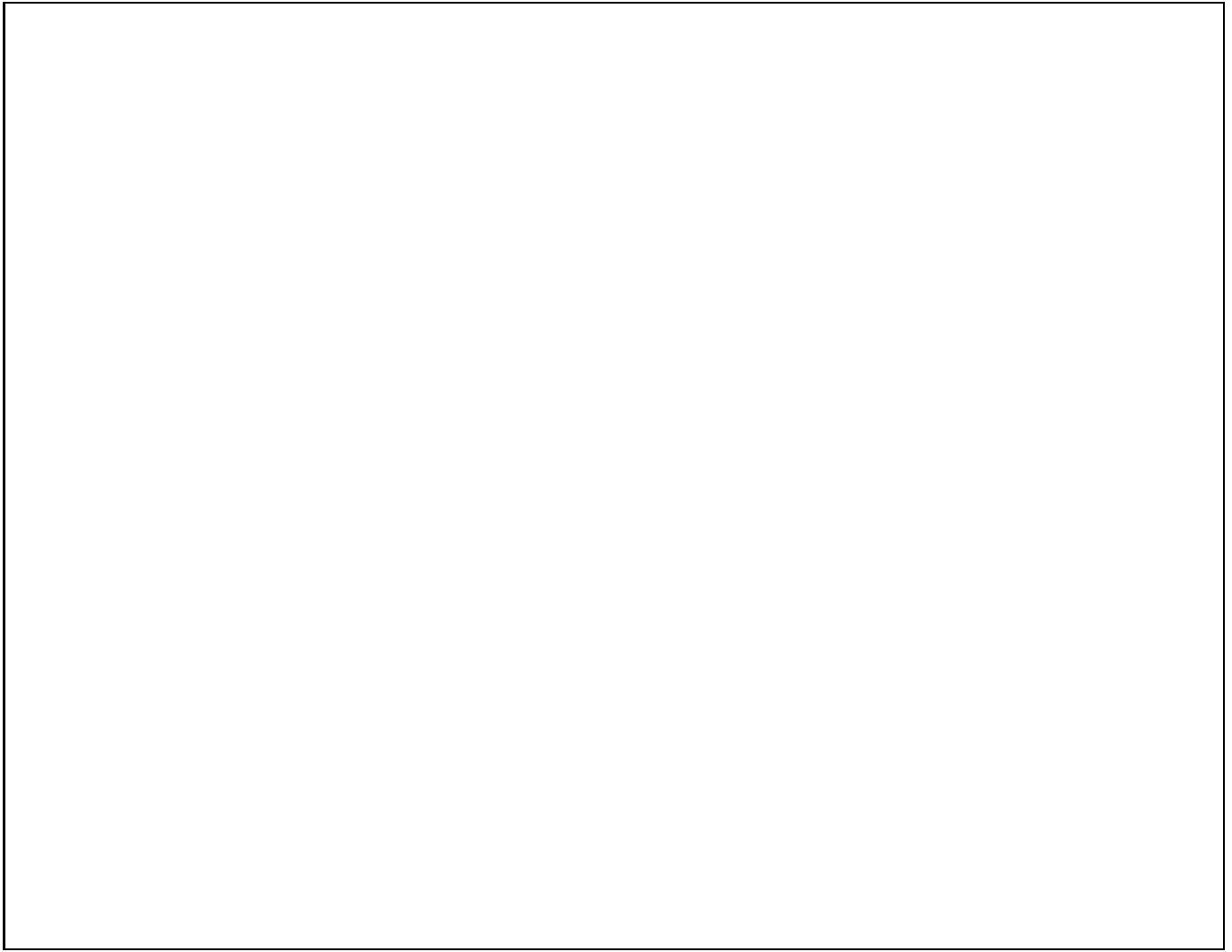
\figure{2b}{Optical spectra of Q0256--0000, Q0301--0035, PKS 0637--75,
MC 1104+16}
\endfig

The journal of the optical spectroscopic observations carried out at the
La Silla Observatory is given in Table 2.

For the reduction process the standard MIDAS facilities
available at the Padova Department of Astronomy and at ESO Garching
have been used.
The raw data were sky-subtracted and corrected for pixel-to-pixel sensitivity
variations by division by a suitably normalized exposure of the spectrum of an
incandescent source. Wavelength calibration was carried out by comparison with
exposures of Helium and Argon lamps. Absolute flux calibration was finally
achieved by observations of standard stars listed by Oke (1974) and Stone
(1977).

For both PKS 0637--75 and MC 1104+16 the optical spectrophotometry has been
obtained quasi-simultaneously with respect to the IUE observations in order
to minimize the effects of variability.
Three spectra of PKS 0637--75 obtained
on December 1986 were normalized to the flux level of Dec 1990
and merged with the 1990 spectrum.

The optical spectra of all the quasars are shown in Fig. 2a-b.

\titlea {Measuring the average flux decrement}

\begfigwid 14.9 cm
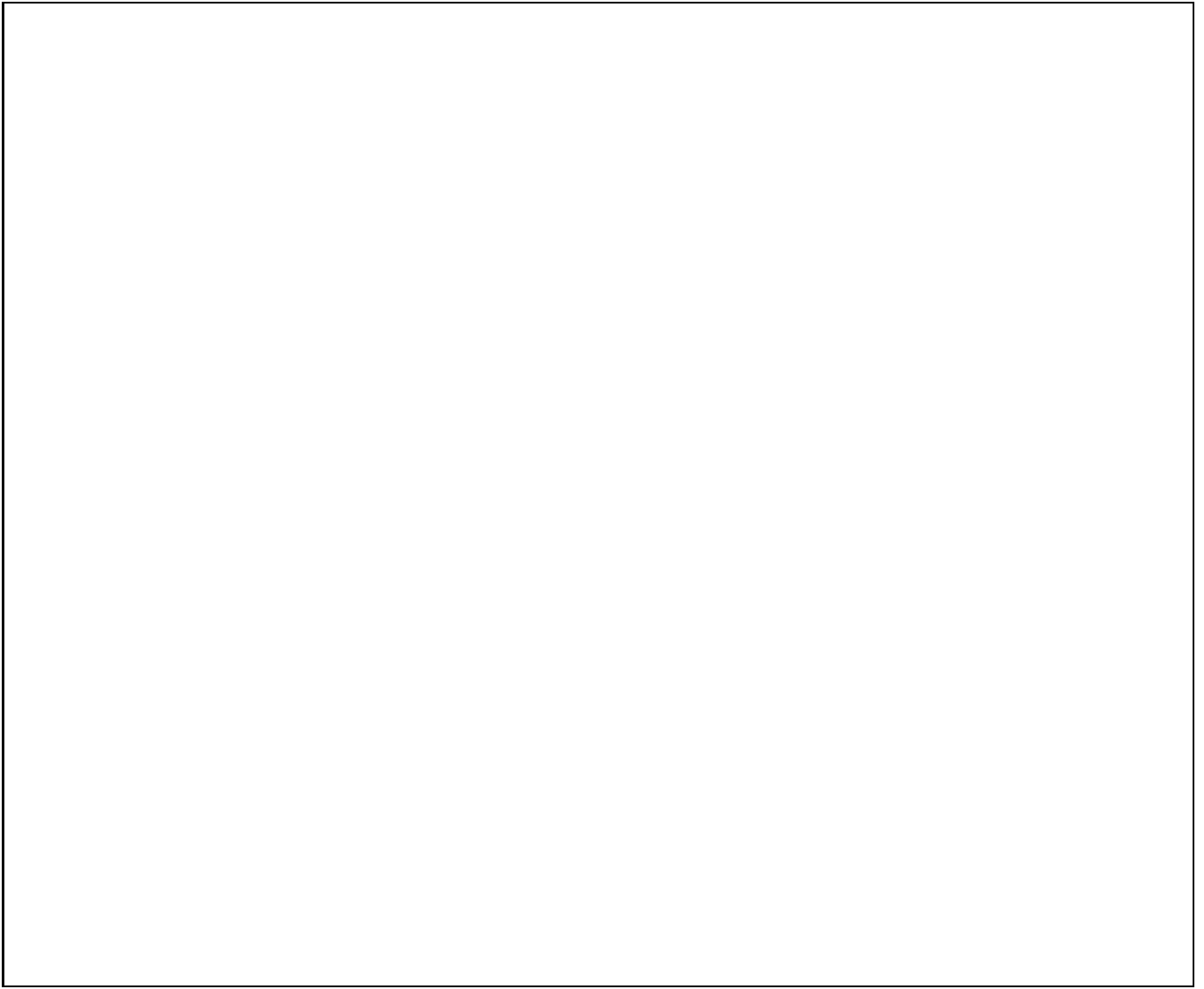
\figure{3}{$log (\nu) - log (F_{\nu})$ combined spectra of PKS 0637--75 and
MC1104+16.}
\endfig

All the spectra have been corrected for galactic
extinction according to Burstein \& Heiles (1982) and rebinned to the QSO rest
frame on the basis of their emission redshift, as shown in Fig. 3,
in a $log (\nu) - log (F_{\nu})$ form.

The slope of the continuum has been
estimated in the form of a power-law $f_{\nu} \propto \nu^{-\alpha}$ (the
dashed line in Fig. 3) on the basis of the ``true continuum'' regions at lower
frequencies with respect to the Ly$_{\alpha}$ emission,
and extrapolated to the regions $A$ and $B$, as defined by Oke and Korycansky
(1982).
The average flux decrements $D_A$ and $D_B$ have
been computed according to the equation:

$$ {D_i = < 1 - f_{obs}/f_{int} > , ~~~~~~ i=A,B ~~~~~~~~~(1)} $$

\noindent where f$_{obs}$ and f$_{int}$ are the observed and ``intrinsic''
fluxes per unit wavelength in the QSO rest frame.

As emphasized in Giallongo and Cristiani (1990), the evaluation of the flux
decrements $D_A$ and $D_B$ is subject to many sources of uncertainty, closely
related to the difficult estimate of the ``intrinsic continuum''. At low
redshifts these problems have relevant consequences on the interpretation of
the data since, as shown below, the derived uncertainties (contrary to the
situation at high redshift) are of the same magnitude as the measured effect.

Composite quasar spectra (Cristiani and Vio 1990, Francis {\it et al.}\ 1991),
used by us as a guide in the procedure of extrapolation,
illustrate how few regions can be considered as indicators of
the ``true'' continuum. In particular the excess emission around C III]
(the blended Fe II 2000 \AA\ complex) is
likely to introduce a bias in the measurements for high redshift QSOs, in
the sense of a steeper estimated continuum and a consequent underestimate
of the Lyman absorption. Our low-redshift measurements, based on
IUE+optical data, could therefore be more accurate but not homogeneous with
respect to the high-redshift data, since the estimate of the ``intrinsic
continuum'' is carried out on different rest-frame intervals.
On the other hand, such a concern is probably less important than it may appear
because a) in the present case, the only effect of the use of the optical+IUE
data with respect to the IUE data only is to restrict the range of
the possible continuum slopes to the steeper ones;
b) the quasars used in the $D_{A,B}$ measurements have redshifts
typically larger than those of the quasars used to construct the composite
spectra, and for higher redshift QSOs the contribution of the Fe 2000 \AA\
complex is probably less important (e.g. Giallongo, Cristiani, Trevese 1992).

To get an idea of the effects due to differences in the spectral ranges
and/or resolutions it is interesting to note that
the continuum slope estimated by us for
PKS 0637--75 ($\alpha = 0.30\pm 0.05$) is significantly flatter than the
value obtained by  O'Brien {\it et al.} (1988), $0.53\pm 0.02$, on the basis of
IUE spectra
and optical photometry, and that for MC1104+16 the spectral index obtained
by Oke {\it et al.} (1984), $0.24\pm 0.1$, on the basis of very low resolution
optical spectrophotometry, is significantly flatter than ours, $0.48\pm0.05$.
The above mentioned indices can also be compared with the ``color-determined''
values of Yu (1987) of 0.84 and 0.60 respectively.

Other non-negligible sources of uncertainty are the matching between the
optical and IUE flux calibrations and, especially in the case of PKS 0637--75,
the amount of galactic extinction adopted.

Since systematic errors are extremely difficult to evaluate, in Table 3 we have
indicated as uncertainties in the $\alpha$, $D_A$, $D_B$ values only the
``internal'' uncertainties deriving from the different possible choices of the
continuum slope and normalization.

\begtabfullwid
\tabcap{3}{Summary of the relevant data}
\halign{ # \hfill & \hfill # \hfill && \hfill #  \hfill \cr
\noalign{\medskip\hrule\medskip}
Name & z$_{em}$ & E$_{B-V}$ & $\alpha_{\nu}$ & $D_A$ & $D_A$ & $D_B$ &
$D_B$ & i.s. Mg II EW \cr
& & (mag) & ($\lambda > 1250$ \AA) & (1050-1170 \AA) & (1040-1190 \AA) &
(~920-1015 \AA) & (~912-1020 \AA) & (\AA) \cr
\noalign{\medskip\hrule\medskip}
PKS 0637--75 & $0.654\pm0.003$ & 0.12 & $0.30\pm0.05$ & $0.07\pm0.04$ &
$0.06\pm0.04$ & $0.18\pm0.05$ & $0.21\pm0.05$ & $6.5\pm0.5$ \cr
MC 1104+16 & $0.632 \pm 0.002$   & 0.02 & $0.48\pm0.05$ &
$0.04~^{+~0.02}_{-~0.01}$ & $0.02~^{+~0.02}_{-~0.01}$ &
$0.11~^{+~0.03}_{-~0.02}$ & $0.10~^{+~0.03}_{-~0.02}$ & $4.0\pm0.3$ \cr
Q0053--4029 & $2.758\pm0.005$ & 0.01 & $0.70\pm0.08$ & $0.18\pm0.04$ &
$0.18\pm0.04$\cr
Q0126--4050 & $2.594\pm0.005$ & 0.01 & $0.35\pm0.10$ & $0.23\pm0.03$ &
$0.19\pm0.03$\cr
Q0249+0222  & $2.805\pm0.005$ & 0.06 & $0.82\pm0.08$ & $0.17\pm0.03$ &
$0.14\pm0.03$\cr
Q0254--0137 & $2.684\pm0.005$ & 0.04 & $0.53\pm0.05$ & $0.20\pm0.02$ &
$0.18\pm0.02$\cr
Q0256--0000 & $3.367\pm0.004$ & 0.06 & $0.61\pm0.05$ & $0.26\pm0.02$ &
$0.22\pm0.02$\cr
Q0301--0035 & $3.205\pm0.005$ & 0.06 & $0.54\pm0.05$ & $0.30\pm0.03$ &
$0.27\pm0.03$\cr
\noalign{\medskip\hrule\medskip}
\noalign{\medskip\hrule\medskip}}\endtab

\titlea {Comparison with models and conclusions}

We have compared the measured average depressions $D_A$ with the
predictions derived on the basis of the model statistics of the
Ly$_{\alpha}$ absorptions superimposed to a synthetic QSO spectrum as in
Giallongo, Gratton \& Trevese (1990) and Giallongo \& Cristiani (1990).

As in the previous papers we have considered the following classes of
absorbers.

(i) The Ly$_{\alpha}$ lines not associated with metal line systems.
These systems evolve in redshift with an average distribution of restframe
equivalent widths according to the following equation:

$$ {\partial ^2 {N(z,W)} \over \partial z \partial W} =
{A \over W^*}(1+z)^{\gamma} \exp{(-W/W^*) ~~~~~~~~~(2)}  $$

\noindent A recent estimate of the Ly$_{\alpha}$ line distribution in the range
$1.8 <z< 3.8$ and $0.2<W<0.5$ ~\AA~ has been obtained from a small
high resolution sample by Giallongo (1991). On average
$W^* =0.2$ ~\AA~ and $\gamma =2.1$ was found.
For $W>0.5$ ~\AA~ a flatter distribution
with $W^*=0.3$ ~\AA~ has been assumed following the equivalent width
distributions obtained from large intermediate resolution samples.
A set of random absorption redshifts is extracted according to equation
(2) with $\gamma =2.1$. We then associate to each redshift a random
equivalent width extracted from a probability distribution
$P(W)\propto \exp{(-W/W^*)}$ using the above $W^*$ values in the
appropriate $W$ ranges.

(ii) Ly$_{\alpha}$ lines associated with metal line systems. We adopt the same
$z,W$ distributions with $\gamma=0.68$
(Sargent, Steidel \& Boksenberg 1989) and $W^*=1.05$ ~\AA~
(Sargent et al. 1980) as in the previous papers.

(iii) Lyman $\beta$, $\gamma$, $\delta$ are computed for each Ly$_{\alpha}$
using a standard curve of growth derived from a Voigt profile. We have
adopted a Doppler parameter value of $b=20$ Km s$^{-1}$ for weak lines
($W<0.5$ ~\AA) and $b=40$ Km s$^{-1}$ for strong lines ($W>0.5$ ~\AA).

(iv) We have also included in the line statistics a population of
damped systems with $W^*\simeq 10$ ~\AA~ to account for the average absorption
observed at very high redshift ($z>4$), (see Giallongo \& Cristiani 1990
for details).
For each emission redshift we compute 100 synthetic QSO absorption
spectra using Voigt profiles with natural damping only.

The available data and simulations at various $z_{em}$ for the average
depressions $D_A$ are shown in Fig. 4. For $z_{em}>2.5$ individual
$D_A$ values are computed from optical spectra of QSOs taken by various
authors at different resolutions (6--100 ~\AA). For $z_{em}<2$
average depressions are derived from IUE spectra.

\begfigwid 13 cm
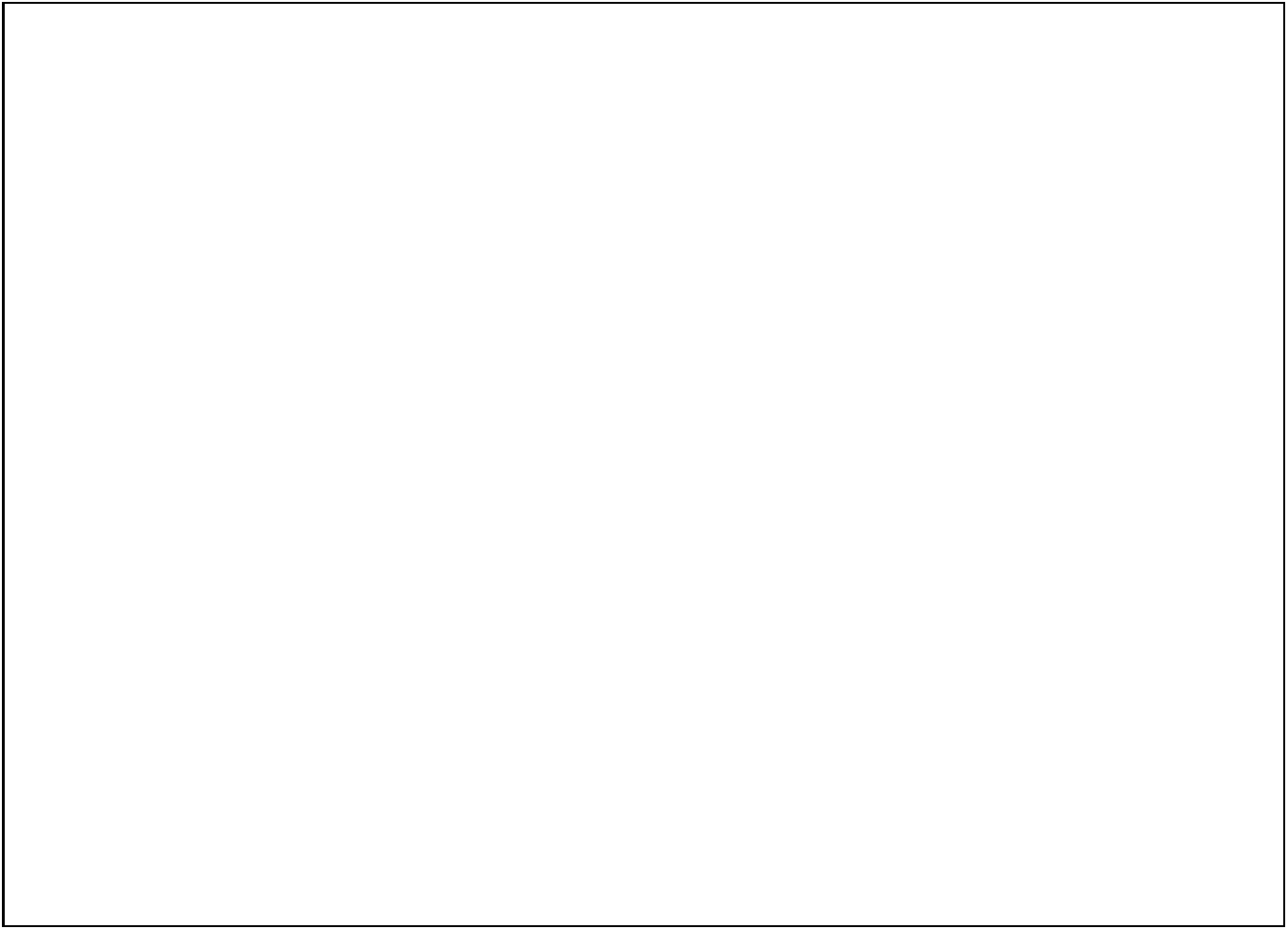
\figure{4}{Comparison between the observations and the simulations
for the average $D_A$ depressions.
Crosses represent the average values from the Monte Carlo simulations.
The contribution of the interstellar absorption lines has been subtracted
to the $D_A$ values observed for PKS 0637--75 and MC 1104+16.}
\endfig

We have represented both the data of O'Brien, Gondhalekar \& Wilson (1988)
and Bechtold et al. (1984). We note that the values derived by
Bechtold et al. are systematically higher, probably because of the
low resolution adopted in the optical spectra (80 ~\AA) which  could
cause an overestimate of the continuum level at wavelengths longer than
the Ly$_{\alpha}$ emission.

When comparing observed and predicted $D_{A,B}$ values on the basis
of the Ly$_{\alpha}$ statistics, we have to remove the contribution
of the interstellar absorption lines, that is non-negligible when the average
depressions are very low (e.g. $<0.1$).
For example, in the high resolution spectrum of 3C 273 (Bahcall {\it et al.}
1991), we can evaluate a value of $D_A\simeq W_{tot}/\Delta \lambda=0.02$ due
to the interstellar absorption lines. We can estimate the analogous
contribution for PKS 0637--75 and MC 1104+16 by considering the interstellar
absorptions observed in the 3C 273 spectrum in the range corresponding to the A
region of PKS 0637--75 and MC 1104+16, and scaling their equivalent widths
according to the ratio between the EW of the 2798 \AA\ Mg II absorption feature
observed in our quasars and the one in 3C 273. In this way an extra $D_A$ value
of 0.009 is derived for PKS 0637--75 and of 0.006 for MC 1104+16.

After this correction we have, at $z\simeq0.65$, an average observed value
for the two QSOs, $D_A = 0.047\pm0.013$ which is larger than but still
consistent, within the relatively large uncertainties, with the one predicted
by the redshift evolution of the Ly$_{\alpha}$ lines $D_A=0.026\pm0.015$.
In this way, marginal
evidence is found for a departure from the standard Lyman $\alpha$ evolution
down to $z\simeq 0.65$. The excess of absorption could be due to the presence
of the diffuse component of neutral hydrogen (the Gunn-Peterson effect)
or to a slower evolution towards low redshifts of the Ly$\alpha$ lines.

Indeed recent HST high resolution spectra obtained at lower redshifts
$z<0.3$ seem to imply a slower evolution with $\gamma \simeq 0.7$
(Bahcall et al. 1992). Adopting this evolutionary rate from $z\simeq 2.5$
down to $z=0.65$ we obtain $D_A=0.047\pm 0.015$ in agreement with our
observed absorptions. Thus no significant Gunn-Peterson effect is detectable
in the interval $0.65<z<4.2$.

The implications of the change in the line statistics would be important for
the physical models of
the Ly$_{\alpha}$ clouds and for the cosmological evolution of the
intergalactic medium and the UV ionizing flux. In the
standard scenario of pressure confined cloud model,
the change in the cosmological evolution of the clouds is related to a quick
drop of the UV ionizing flux at very low redshifts (Ikeuchi \& Turner 1991).

The uncertainties involved in the process of estimating the continuum slope
and normalization, matching the spectra obtained in different spectral ranges
with different instruments and detectors, evaluating correctly the galactic
extinction are such to indicate that an improvement of the present estimates
at low redshifts with low-resolution spectrophotometry will be extremely
difficult. Higher resolution line statistics at low z is obviously the way
to follow.

\acknow {It is a pleasure to thank Paola Andreani for helpful discussions and
suggestions. Fabio La Franca acknowledges the financial support of a
research fellowship of the {\it Fondazione Ing. Aldo Gini}.
}

\begref{References}
\ref Adam, G., 1978, A\&AS {\bf 31}, 151
\ref Bahcall, J. N., Jannuzi, B. T., Schneider, D. P., Hartig, G. F.,
Bohlin, R., Junkkarinen, 1991, ApJ {\bf 377}, L5
\ref Bechtold, J., Green, F., Weymann, R. J., Schmidt, M., Estabrook, F. B.,
Sherman, R. D., Wahlquist, H. D. \& Heckman, T. M., 1984, ApJ {\bf 281}, 76
\ref Bohlin, R. C., \& Grillmair, C. J., 1988, ApJS {\bf 66}, 209
\ref Boyle, B. J., Shanks, T. \& Peterson, B. A., 1988, MNRAS, {\bf 235}, 935
\ref Burstein, D. \& Heiles, C., 1982, AJ {\bf 87}, 1165
\ref Cassatella, A., Ponz, D., Selvelli, P.L. \& Vogel, M.,
1988, {\it ESA IUE Newsletter} {\bf 31}, 1
\ref Cristiani, S. \& Vio, R., 1990, A\&A {\bf 227}, 385
\ref Ellingson, E., Yee, H. K. C. \& Green, R. F., 1991, ApJ, {\bf 371}, 49
\ref Francis, P. J., Hewett, P. C., Foltz, C. B., Chaffee, F. H.,
Weymann, R. J. \& Morris, S. L., 1991, ApJ {\bf 373}, 465
\ref Giallongo, E., 1991, MNRAS {\bf 251}, 541
\ref Giallongo, E., \& Cristiani, S., 1990, MNRAS {\bf 247}, 696
\ref Giallongo, E., Gratton, R. G. \& Trevese, D., 1990, MNRAS, {\bf 244}, 450
\ref Giallongo, E., Cristiani, S., \& Trevese, D., 1992, ApJL, in pres
\ref Holm, A.V., Bohlin, R.C., Cassatella, A., Ponz, D., \& Schiffer
III, F.H., 1982, A\&A  {\bf 112}, 341
\ref Ikeuchi, S., \& Turner, E. L., 1991, ApJ Lett., {\bf 381}, L1
\ref O'Brien, P. T., Gondhalekar, P. M. \& Wilson, R., 1988, MNRAS
{\bf 233}, 801
\ref Oke, J. B., 1974, ApJS, {\bf 27}, 21
\ref Oke, J. B., \& Korycansky, D. G., 1982, ApJ {\bf 255}, 11
\ref Oke, J. B., Shields, G. A. \& Korycansky, D. G., 1984, ApJ {\bf 277}, 64
\ref Sargent, W. L. W., Young, P. J., Boksenberg, A. \& Tytler, D.,
1980, ApJS {\bf 42}, 41
\ref Sargent, W. L. W., Steidel, C.C., \& Boksenberg, A., 1989, ApJS
{\bf 69}, 703
\ref Schneider, D. P., Schmidt, M.,  \& Gunn, J. E., 1989, AJ {\bf 98}, 1507
\ref Schneider, D. P., Schmidt, M.,  \& Gunn, J. E., 1991, AJ {\bf 101}, 2004
\ref Steidel, C. C., \& Sargent, W. L. W., 1987, ApJ, {\bf 313}, 171
\ref Stone, R. P. S., 1977, ApJ {\bf 218}, 767
\ref Teays, T. \& Garhart, M. P., 1990, {\it NASA IUE Newsletter},
{\bf 41}, 94
\ref Yu, K. N., 1987, Ap\&SS {\bf 134}, 35
\bye